\documentclass[aps,prl,reprint,groupedaddress]{revtex4-1}

\usepackage{amsmath}            
\usepackage{amssymb}            
\usepackage{latexsym}
\usepackage{subfigure}
\usepackage{rotating}
\usepackage{graphicx}
\usepackage{float}

\begin{document}
\graphicspath{{Figs/}{Figs/PDF}}

\title{A Microstructural View of Burrowing
with RoboClam} 
\author{K. N. Nordstrom$^{1,3}$, D. S. Dorsch$^{2}$, W. Losert$^{1}$, A. G. Winter, V$^2$}
\affiliation{$^1$Institute for Physical Science and Technology, and Department of Physics, University of Maryland, USA}
\affiliation{$^2$Department of Mechanical Engineering, Massachusetts Institute of Technology, USA}
\affiliation{$^3$Current affiliation: Department of Physics, Mount Holyoke College, USA}

\date{\today}

\begin{abstract}
RoboClam is a burrowing technology inspired by \it Ensis directus\rm, the Atlantic razor clam. Atlantic razor clams should only be strong enough to dig a few centimeters into the soil, yet they burrow to over 70 cm. The animal uses a clever trick to achieve this: by contracting its body, it agitates and locally fluidizes the soil, reducing the drag and energetic cost of burrowing. RoboClam technology, which is based on the digging mechanics of razor clams, may be valuable for subsea applications that could benefit from efficient burrowing, such as anchoring, mine detonation, and cable laying. We directly visualize the movement of soil grains during the contraction of RoboClam, using a novel index-matching technique along with particle tracking. We show that the size of the failure zone around contracting RoboClam, can be theoretically predicted from the substrate and pore fluid properties, provided that the timescale of contraction is sufficiently large. We also show that the nonaffine motions of the grains are a small fraction of the motion within the fluidized zone, affirming the relevance of a continuum model for this system, even though the grain size is comparable to the size of RoboClam. 

\begin{description}
\item[PACS numbers] 87.19.rs, 81.05.Rm, 81.40.Np, 81.70.Bt
\end{description}
\end{abstract}

\maketitle
\section{Introduction}

As we all know from common experience, a bowl of sand will slosh around much like a bowl of soup. But stick your finger into each, and the material resists quite differently. The soup offers almost no resistance, and the sand's resistance increases quickly until you can't push any further. But this is more than just a whimsical exercise; burrowing in granular materials is of great technological interest, in applications such as anchoring vessels and laying undersea communication cables.
 
Many animals also have a vested interest in the manipulation of granular materials, needing to walk, swim, or burrow through them. As such, they have evolved unique locomotion strategies to make their way, often to optimize efficiency \citep{trueman1975}. The sandfish lizard ({\it S. scincus}) swims through sand, with motion resembling the undulations of a fish \cite{maladen2009}. Clam worms ({\it N. virens}) use crack propagation to burrow in mud-like gelatin \cite{dorgan2005}. Nematodes ({\it C. elegans}) move efficiently via reciprocating motion in saturated granular media \cite{wallace1968,jung2010}.

\begin{figure}[h!]
\centering
\includegraphics[width=.9\columnwidth]{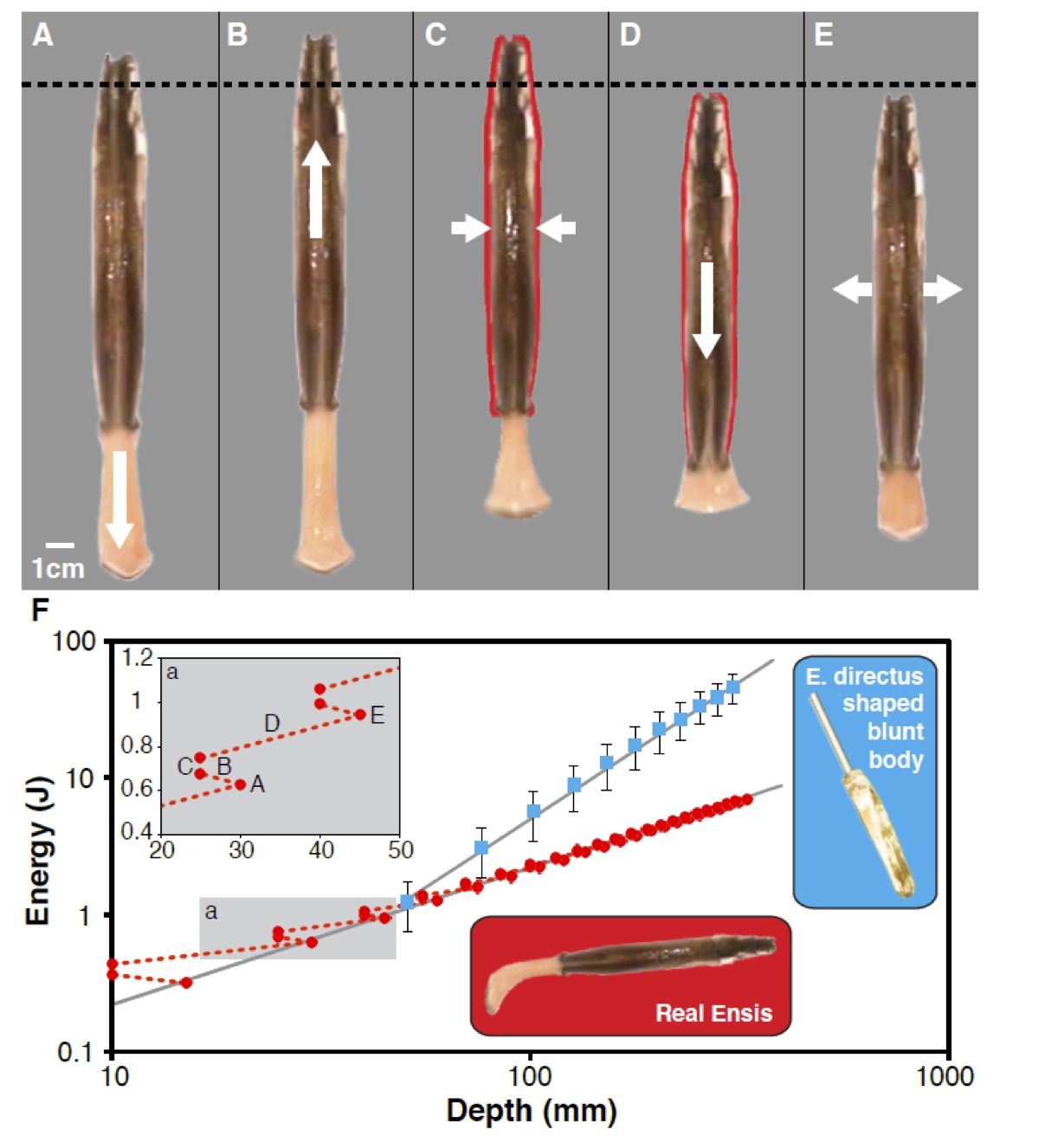}
 \caption{ (A-F) {\it E. directus} digging cycle kinematics and energetics. 
White arrows indicate valve movements. Red silhouette denotes valve geometry in expanded state, before contraction. 
   (A) Extension of foot at initiation of digging cycle.
   (B) Valve uplift.
   (C) Valve contraction, which pushes blood into the foot, expanding it to serve as a terminal anchor.
   (D) Retraction of foot and downwards pull on the valves.
   (E) Valve expansion, reset for next digging cycle.
   (F) Energetic cost to reach burrow depth for {\it E. directus} and a blunt body of the same size and shape as the animal pushed into static soil. } \label{fig:clam}
\end{figure}

In contrast to a liquid, in which viscosity and density do not change with depth, particles within a static granular material experience contact stresses, and thus frictional forces, that scale with the surrounding pressure, resulting in shear strength that increases linearly with depth \cite{terzaghi1996}. This means that submerging devices such as anchors can be costly, as insertion force $F(z)$, increases linearly with depth $z$ \cite{robertson1983}, resulting in an insertion energy, $E=\int F(z)\,  dz$,  that scales with depth squared.

The Atlantic razor clam, {\it Ensis directus}, can produce approximately 10 N of force to pull its valves into soil \cite{trueman1967}.
Using measurements from a blunt body the size and shape of {\it E. directus} pushed into the animal's habitat substrate, this level of force should enable the clam to submerge to approximately 1--2cm \cite{2012expbiorazorclam}. But in reality, razor clams dig to 70cm \cite{holland1977} indicating that the animal must manipulate surrounding soil to reduce burrowing drag and the energy required for submersion.

{\it E. directus} burrows by using a series of valve and foot motions to draw itself into underwater soils (Figs. \ref{fig:clam}A--E). Comparing this performance to the energy required to push an {\it E. directus}-shaped blunt body to burrow depth in the animal's habitat substrate using steady downward force (Fig. \ref{fig:clam}F), we find the animal is able to reduce its required burrowing energy by an order of magnitude, even taking into account energy spent manipulating its valves -- motions that do not directly contribute to downward progress \cite{2012expbiorazorclam}. 

\begin{figure}[t]

\centering


\includegraphics[width=1\columnwidth]{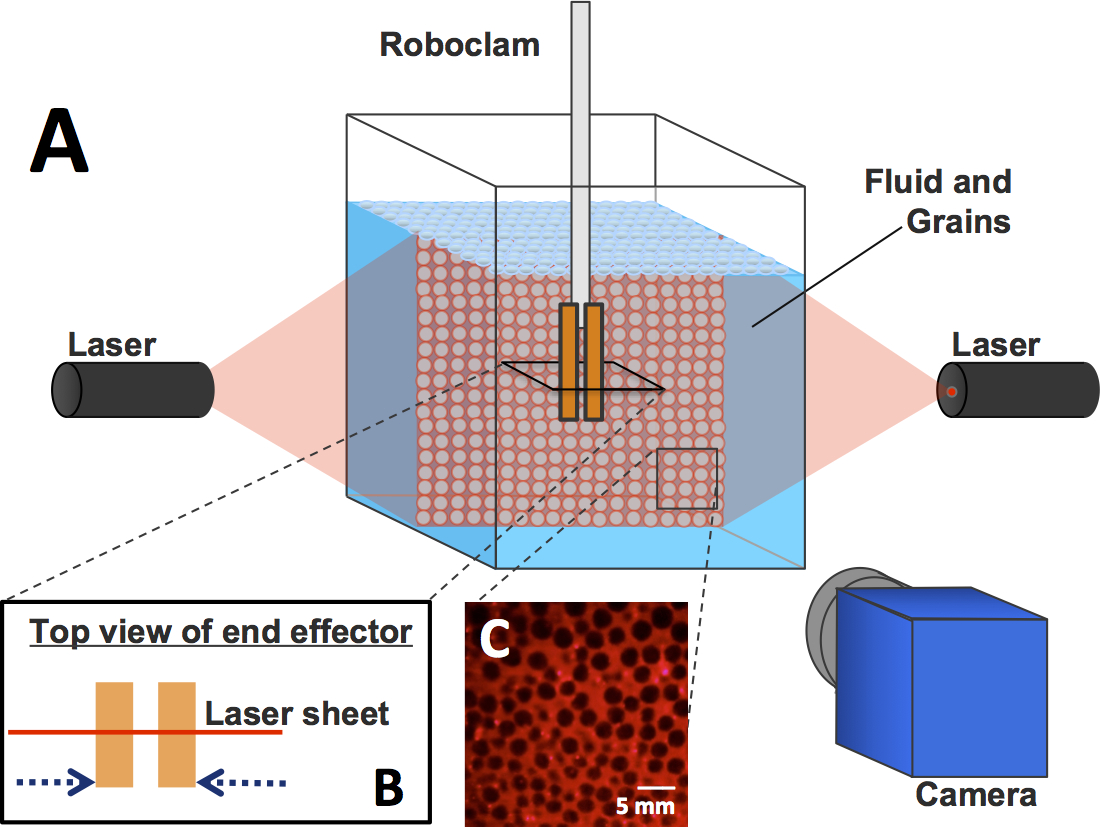}
 \caption{
The experimental setup. A)  The immersion fluid is fluorescent and index-matched to the granular material. A red laser sheet fluorescently excites a slice within the sample and so the camera captures that slice only. The slice recorded is the plane of motion of the end effector. Images are captured at 150 fps as the end effector is contracting. (B) A top view of the end effector's motion, illustrating the plane the camera is capturing. (C) A portion of an image taken. Due to the fluorescent dye within the fluid, the fluid is bright red within the images, and the grains are dark. Images are subjected to particle identification and tracking algorithms, which yield measurements of individual grain trajectories.}
 \label{fig:expt}
\end{figure}

But even though these valve motions do not advance the animal downwards, these motions are critical. The uplift and contraction of {\it E. directus'} valves during burrowing locally agitate the soil (Fig. \ref{fig:clam}B-C) and create a region of fluidization around the animal \cite{2012expbiorazorclam}. Moving through fluidized, rather than static, soil reduces drag forces on the animal to within its strength capabilities \cite{2012expbiorazorclam}. These fluidized substrates can, to first order, be modeled as a generalized Newtonian fluid with depth-independent density and viscosity that are functions of the local packing fraction \cite{einstein1906,frankel1967,krieger1959,eilers1941,ferrini1979,maron1956}. As a result, burrowing via localized fluidization requires energy that scales linearly with depth, rather than depth squared for moving through static soil (Fig. \ref{fig:clam}F).

 {\it E. directus} is an attractive candidate for biomimicry when judged in engineering terms: its body is large (approximately 20 cm long, 3 cm wide); its shell is a rigid enclosure with a one degree of freedom hinge; it can burrow over half a kilometer using the energy in an AA battery \cite{energizer2009}; it can dig quickly, up to 1 cm/s \cite{trueman1967}, and it uses a purely kinematic event to achieve localized fluidization, rather than requiring additional water pumped into the soil. There are numerous industrial applications that could benefit from a compact, low-energy, reversible burrowing system, such as anchoring, subsea cable installation, mine neutralization, and oil recovery. An {\it E. directus}-based anchor should be able to provide more than ten times the anchoring force per insertion energy as existing products \cite{2011jicbroboclam}. 

In previous work we have discussed the performance of RoboClam, an {\it E. directus}-inspired robot \cite{BioMim}. By using a genetic algorithm we found optimal parameters for digging efficiency. These parameters corresponded to specific contraction and expansion times of the robot. These timescales, and the size of the fluidized zone, can be predicted by a model derived from soil, fluid, and solid mechanics theory, and only require input of two commonly measured geotechnical parameters: the coefficient of lateral earth pressure and the friction angle. While the previous optimization testing was fruitful and was consistent with the model in terms of the optimal timescales, what remains is to test the model by directly measuring the size of the fluidized zone. 

In this paper, we use a refractive index-matching technique to directly record the motion of the grains \it within \rm a typically murky 3D granular system, while the device is contracting. We are then able to compare the size of the real fluidized region to that predicted by the model. We vary the contraction timescale, and show under what conditions the model breaks down, an important piece of knowledge for technical development of new devices. 

\section{Experimental Details}

RoboClam replicates the digging kinematics of \it E. directus \rm (Fig. \ref{fig:clam}). Instead of using valves, RoboClam uses a simple mechanical system to actuate the end effector.  The robot consists of 3 main parts: the ``end effector,'' which is two pieces of metal (``shells'') able to diverge or converge horizontally, imitating the valve motion of the organism. The end effector is attached to a hollow extruded rod which is fixed to a platform. Within this rod is a second rod which terminates on either end outside of the hollow rod. At one end it terminates in a wedge inside the end effector, at the other, it terminates in a plunger outside of the hollow rod. 

Moving the plunger up thus moves the wedge up (but does not affect the vertical position of the end effector), which then moves the sides of the end effector inwards. Moving the plunger down has the opposite effect. Thus the inner rod controls the in/out motion of the end effector. The outer rod can itself be moved to control the up/down motion of the end effector \cite{BioMim} but we will not consider this complication here. For these experiments, we solely focus on the contraction of the effector, which acts to fluidize the surrounding soil. We move this plunger with a stepper motor to control the contraction time. 

The end effector is of similar size as a juvenile \it E. directus \rm (9.97 cm long and 1.52 cm wide). It also has the capability to contract up to 6.4 mm, which is about twice the contraction ability of the adult organism. This enhanced capability was added in order to test the effects of greater movements in the artificial system. The end effector is sealed within a neoprene boot to prevent particles from jamming the valve expansion/contraction. Further design details and testing results can be found in \cite{BioMim} and \cite{winter2010thesis}.

\begin{figure}[b]
\centering
\includegraphics[width=1\columnwidth]{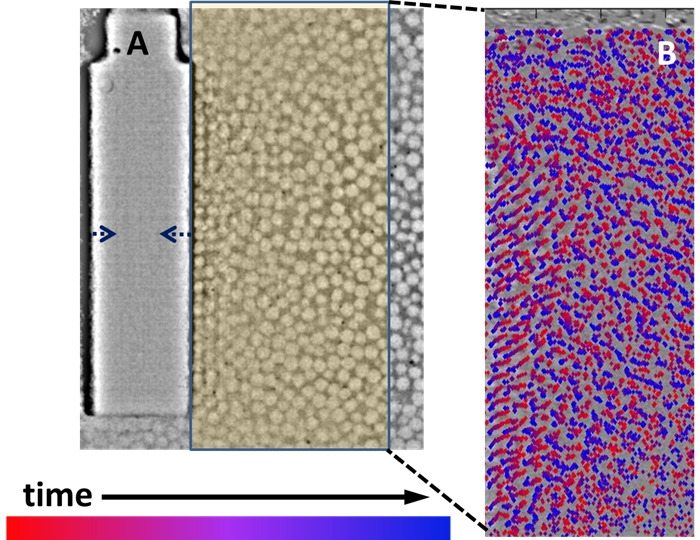}
 \caption{(A) An actual sample image (inverted) from an experiment. The blue arrows indicate the contraction motion of the end effector, also shown. The yellow box illustrates the area used for data analysis, to the right of the contracting end effector. (B) Particle tracks for this same experiment. Particles are identified and tracks are made by connecting particles between frames. The overlaid points are particle positions for a full contraction cycle, and the color indicates the relative time in the cycle. A streak of red-to-blue represents a unique particle track. }
 \label{fig:tracks}
\end{figure}

Our experimental setup is shown in (Fig. \ref{fig:expt}A). In order to transcend the ``clear as mud" nature of granular materials, we used an index-matching technique that allows us to see inside a normally opaque sample. Our grains are 3 mm glass borosilicate spheres (Glen Mills). They are poured to fill a clear box,  15 cm on each side.  The box is then filled with a mixture of DMSO (about 95 percent by weight), 0.12 M hydrochloric acid, and Nile Blue 690 perchlorate dye (trace). The fluid mixture is tuned to match the index of refraction of the grains, so that the index mismatch is less than 0.005. A laser sheet is set to illuminate a plane which captures the contraction motion (Fig. \ref{fig:expt}B), resulting in bright fluid and dark grains (Fig. \ref{fig:expt}C) \cite{Joshua}. We image this slice during RoboClam's contraction using a high-speed, light-sensitive PCO.edge camera (PCO AG), taking video data at speeds up to 150 fps. 

From these videos we can extract the positions of the grains at all times (Fig. \ref{fig:tracks}B) during a contraction of the Clam using established particle tracking routines \cite{Steve}. As the system is symmetric, we focus only on grains directly to the right of the contraction. From these positions, we are able to calculate particle displacements, local void fractions, and nonaffine motions of the grains \cite{Nordstrom}. We measure these quantities throughout the contraction. We further explore the phase space of this system by varying the contraction speed of RoboClam over  an order of magnitude, corresponding to inward contraction times of $0.053<t_{in}<0.378$ s. (The contraction speed of the organism is approximately 0.2 s \cite{trueman1967}).

We have measured the grain motions in the plane perpendicular to the main contraction motion. We find no substantial motion in this plane, which indicates the response of the grains is almost solely in the direction of contraction. We have also measured the motion in two planes parallel to the motion, but away (1.5 cm and 4.5 cm) from the edge of the end effector. We find no substantial grain motion in these fields of view. Both observations indicate solely measuring motion in the central plane is sufficient -- out of plane motion is insubstantial. The rest of this paper will discuss the mechanics within this central plane. 

\section{Modeling The System}

We start by briefly reviewing the results found for {\it E. directus} \cite{2012expbiorazorclam}. As {\it E. directus} contracts, it reduces the level of stress acting between its sides and the surrounding soil. As the sides were (in effect) supporting the soil, this causes a stress imbalance. When {\it E. directus} initiates contraction, rather the stress imbalance creates a zone of active failure, specifically creating a failure wedge determined by the friction angle of the soil. The discontinuity in the failure surface enables  the fluidization: particles inside the failure zone are free to move once the clam contracts, while those outside it are stuck in a static pile. The motion of the clam reduces the volume of the animal, which draws pore fluid towards the animal. This creates a locally fluidized region of lower packing in the granular material. The particles free to move are then advected by the pore fluid, which moves inward with {\it E. directus} . The failure wedge is of utmost importance here; without the wedge, all particles will follow the movement of the fluid, and effectively not create a special fluidized zone.  


To test whether these results are also applicable to RoboClam, we start by looking into the fluid dynamics of the system. Assuming Stokes drag (as was shown to be applicable to \it E. directus \rm \cite{2012expbiorazorclam}), the critical time required for a soil particle to reach the pore fluid velocity can be estimated through conservation of momentum:

\begin{equation}
m_p \frac{d v_p}{d t} = 6 \pi \mu_f D (v_v - v_p) \to t_{crit} = \frac{{D}^2\rho_p}{36\mu_f}, 
   \label{eqn:advectiontime}
\end{equation}

\noindent where $D$ is the diameter of a particle, $\mu_f$ is the viscosity of the fluid, $m_p$ is the mass of a particle, $\frac{d v_p}{d t}$ is the acceleration of a particle, and $\rho_p$ is the density of a particle. For the 3 mm borosilicate glass beads in DMSO used in our experiments, $t_{crit} = 0.275$ s. 

In our experiments, we vary the inward contraction time, $t_{in}$. For some experiments $t_{crit}<t_{in}$ and vice versa for others. When $t_{crit}$ is less than the contraction time, the particles can be considered inertialess \cite{2012expbiorazorclam} and are advected with the pore fluid during contraction. When it is greater, we posit that particles will be less able to advect with the flow because of their inertia, resulting in slower particles within the fluidized region and a smaller fluidized region. (As we vary the timescale an order of magnitude only, we do not expect to enter a fast contraction regime where none or few particles are advected.)

\begin{figure}[ht!]
\centering
\includegraphics[width=1\columnwidth]{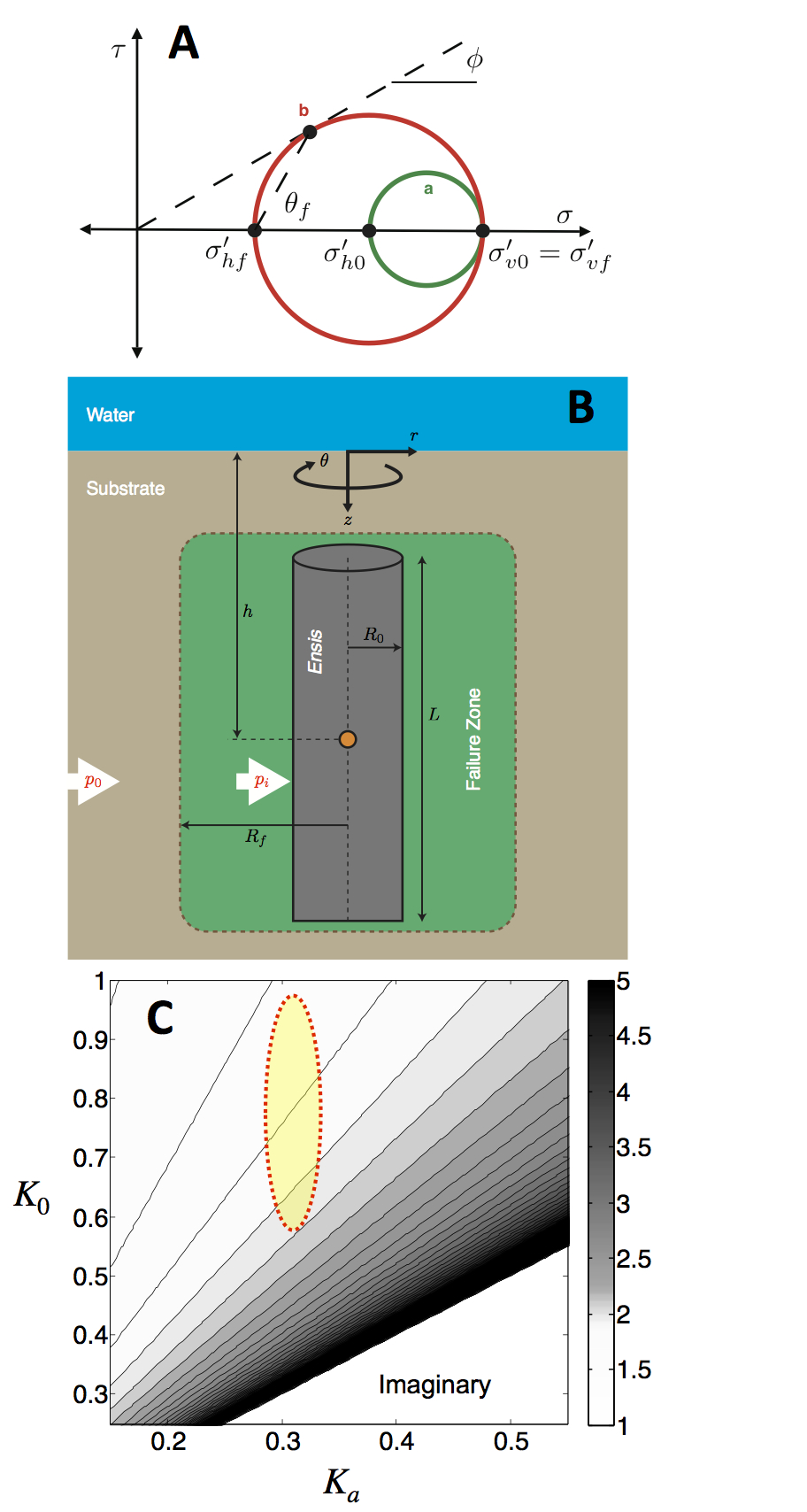}
 \caption{Details of the model. (A) Mohr's circles of stress states for equilibrium (a) and active failure (b). (B) Cylindrical model of soil failure around the  contracting end effector. As
RoboClam contracts it reduces the pressure acting between its body and the
soil, $p_i$, below that of the equilibrium lateral soil pressure, $p_0$. This stress imbalance
induces a localized failure zone around the animal.$R_0$ is  RoboClam's expanded size, and $R_f$ is the size of the failure zone.(C) Predicted size of the failure zone around the end effector, using the full range of possible values for $K_0\in[0.31,1.0]$ and $K_a\in[0.19,0.52]$. The yellow oval corresponds to the expectation for our experimental system.}
 \label{fig:model}
\end{figure}

However, there is another effect competing with fluidization. As the soil begins to fail, it will tend to naturally landslide downward at a failure angle $\theta_f$. At this point, the shear stresses in the soil are equal to its shear strength. This condition is shown in Fig. \ref{fig:model}A, with the applied stress circle b tangent to the failure envelope, which lies at the same angle as the friction angle of the soil $\varphi$, a property commonly measured during a geotechnical survey. The failure angle is the transformation angle between the principle stress state and the stress state at failure. This angle can also be determined by connecting the tangency point on the failure envelope, the horizontal effective stress at failure $\sigma^\prime_{hf}$, and the principle stress axis (Fig. \ref{fig:model}A), and is given by
 
\begin{equation}
\theta_f= \frac{\pi}{4}+\frac{\varphi}{2}.
   \label{eqn:failureangle}
\end{equation}

\begin{figure*}[ht!]
\centering
\includegraphics[width=2\columnwidth]{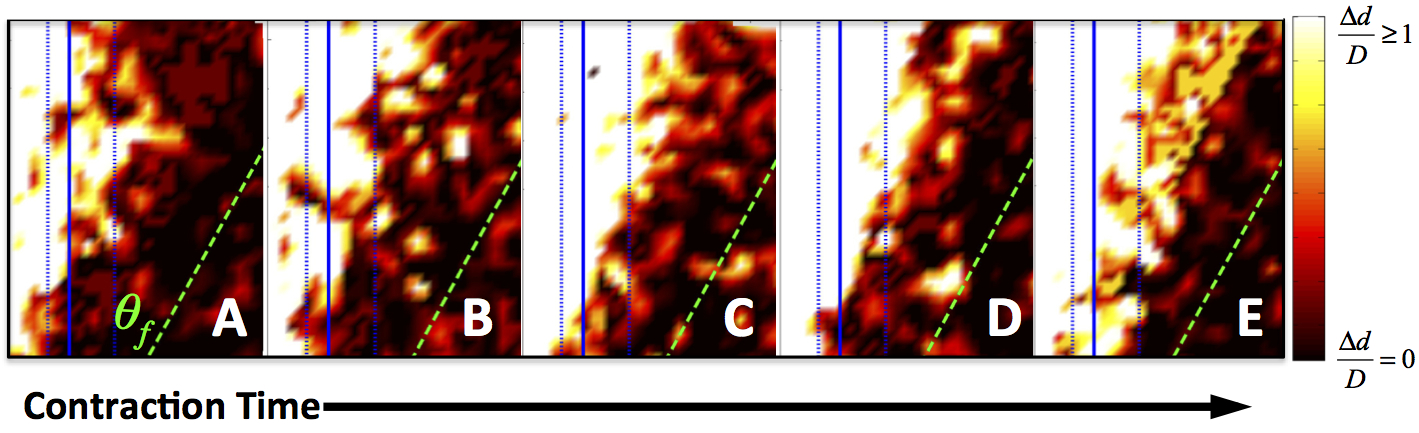}
 \caption{Contour plots of the substrate's local speed for different contraction times, after the contraction. In all images the end effector is to the left of the image and is not shown. The shortest contraction times are on the left, and specifically:  (A) $t_{in}=0.053$ s, (B) $t_{in}=0.153$ s, (C) $t_{in}=0.247$ s, (D) $t_{in}=0.295$ s, and (E) $t_{in}=0.378$ s. The colorscale is the same in all plots, and scaled such that active regions moving displacing more than the cutoff distance $D$ are in white, and static regions are in black. The angle of the failure wedge is predicted by the green dashed line. The blue solid line shows the prediction for the size of the fluidized zone, with the dashed blue lines representing the range of possible sizes.}
 
\label{fig:displacement}
\end{figure*}

\noindent Equation \ref{eqn:failureangle} was used to plot the failure angle in Fig. \ref{fig:displacement}, with the friction angle of the substrate measured as $34^\circ$.

For digging efficiency, the creation of the fluidized zone must occur at a faster timescale than that required for the soil to naturally fail and landslide towards the end effector. In our material, the landslide \cite{winter2010thesis} time is approximately 0.5 s, which is comparable to but greater than our largest full contraction period. Thus we are not generally competing with landslide effects. Further, we note that while  this is an important design consideration for efficient digging, this does not alter the size of the fluidized zone, which is our main scope.


Another competing factor to fluidization is the sedimentation of the particles themselves. If the particles settle on a faster timescale than the contraction, fluidization will not occur. Using the Richardson-Zaki equation ($v_s=v_t\epsilon^n$, where $v_t$ is the terninal velocity of a single particle in a fluid, $\phi$ is the void fraction, and $n$ is the settling index $\approx$ 4.8 \cite{2012expbiorazorclam} to estimate the particle settling time, we find that this is about 3 s, and so not a concern for this experiment, however, it certainly could be important for design considerations. More realistic soils, i.e. smaller particles, will in general have even larger settling times. 

Figure \ref{fig:model}A shows a Mohr's circle representation \cite{hibbeler2000mohr} of the effective stress states at equilibrium, before contraction (circle a), and during the initiation of contraction, which brings the soil into an active failure state, by an imbalance between radial and vertical stresses (circle b). Effective stress is the actual stress acting between soil particles, neglecting hydrostatic pressure of the pore fluid, and is denoted in this paper with a prime. The term ``active" corresponds to the reduction (rather than increase) of one of the principal stresses to induce failure \cite{terzaghi1996}.

To describe the size of the fluidized zone, we turn to a model of RoboClam as a cylinder with contracting radius that is embedded in saturated soil (Fig. \ref{fig:model}B). To neglect end effects, the length of the cylinder is considered to be much larger than its radius. The relaxation in pressure can be considered quasi-static and elastic \cite{terzaghi1996}. The radial and hoop stress distribution in the substrate can be described with the following thick-walled pressure vessel equations \cite{timoshenko1970vessel}, which have been modified to geotechnical conventions (with compressive stresses positive) and to reflect an infinite soil bed in lateral directions \cite{BioMim, winter2010thesis}. Due to the radial symmetry of this model, this will also work for our system: the center plane of each system will be identical. 

\begin{eqnarray}
\sigma_r &=& \frac{R_0^2 (p_i - p_0)}{r^2} + p_0
   \label{eqn:clamstressr} \\
\sigma_\theta &=& -\frac{R_0^2 (p_i - p_0)}{r^2} + p_0,
   \label{eqn:clamstresstheta}
\end{eqnarray}

\noindent where $\sigma_r$ is total radial stress, $\sigma_\theta$ is total hoop stress, $R_0$ is RoboClam's size before contraction, $p_i$ is the pressure acting on Roboclam, and $p_0$ is the natural lateral equilibrium pressure in the soil. It is important to note that these equations still hold if there is a body force acting in the $z$-direction, such as in soil. In this case, the pressure vessel equations describe the state of stress within annular differential elements stacked in the $z$-direction. The total vertical stress is given as

\begin{equation}
\sigma_z = \rho_t g h,
   \label{eqn:clamstressz}
\end{equation}

\noindent where $h$ is the clam's depth beneath the surface of the soil, $\rho_t$ is the total density of the substrate (including solids and fluids), and $g$ is the gravitational constant. It should be noted that there are no shear stresses within the soil in principal orientation, as $\tau_{r z} = \tau_{\theta z} = 0$ because RoboClam is modeled with a high aspect ratio ($L \gg R_0$) and $\tau_{r \theta} = 0$ because of symmetrical radial contraction.

The undisturbed horizontal effective stress in the substrate is determined by subtracting hydrostatic pore pressure $u$  from the natural lateral equilibrium pressure:

\begin{equation}
\sigma'_{h0} = p_0-u.
   \label{eqn:horizeffstress}
\end{equation}

\noindent The undisturbed horizontal and vertical effective stresses can be correlated through the coefficient of lateral earth pressure:

\begin{equation}
K_0 = \frac{\sigma'_{h0} }{\sigma'_{v0}},
   \label{eqn:coefflatearth}
\end{equation}

\noindent which is a soil property that can be measured through geotechnical surveys \cite{terzaghi1996,lambe1969}. By also knowing the void fraction of the soil  $\phi$ and the particle and fluid density, $\rho_p$ and $\rho_f$ respectively, $p_0$ can be determined as 
\begin{equation}
p_0 = K_0 \sigma'_{v0} + u = K_0 g h (1 - \phi) (\rho_p - \rho_f) + \rho_f g h.
   \label{eqn:natsoilpress}
\end{equation}

Failure of the substrate will occur when $p_i$ is lowered to a point where the imbalance of two principle effective stresses produces a resolved shear stress that exceeds the shear strength of the soil. This resolved failure shear stress can be created by an imbalance between radial and vertical stresses (Fig. \ref{fig:model}A, circle b) or radial and hoop stresses. In real systems, the radial-hoop mode [cite biomim] is dominated by the radial-vertical mode. Further, experimentally, we are in a quasi-2D realization of this model, and so radial-vertical modes only truly apply.  

The relationship between stresses at active failure (circle b) is: 

\begin{equation}
\frac{\sigma'_{rf}}{\sigma'_{vf}} = \frac{\sigma'_{rf}}{\sigma'_{\theta f}} = \frac{1 - \sin \varphi}{1 + \sin \varphi} = K_a,
   \label{eqn:coeffactivefail}
\end{equation}

\noindent where the subscript $f$ denotes the stresses at failure and $K_a$ is referred to as the coefficient of active failure. 

Soil failure due to an imbalance between radial and vertical stresses will occur when the applied radial effective stress equals the radial stress at failure. The radial location of the failure surface in this condition, $R_{f_{rv}}$, can be found by combing Eq. \ref{eqn:clamstressr} for radial stress with Eqs. \ref{eqn:horizeffstress}, \ref{eqn:coefflatearth}, and \ref{eqn:coeffactivefail}, and realizing that the vertical effective stress at failure and equilibrium is unchanged, namely  

\begin{eqnarray}
 \sigma'_r \big|_{r = R_{f_{rv}}} &=& \sigma'_{rf}  \nonumber\\
\frac{R_0^2 (p_i - p_0)}{R_{f_{rv}}^2} + p_0 - u &=& \frac{K_a}{K_0} (p_0 - u) \nonumber
\end{eqnarray}

\noindent yielding the dimensionless radius for radial-vertical stress imbalance-induced failure:

\begin{equation}
\frac{R_{f_{rv}}}{R_0} = \left[ \frac{p_i - p_0}{\left(\frac{K_a}{K_0} - 1 \right) (p_0 - u)} \right]^\frac{1}{2}.
   \label{eqn:failradvert}
\end{equation}

We assume $p_i$ is about zero, corresponding to complete stress release between RoboClam's sides and the surrounding soil, and $u \approx 0.5 p_0$ because of the relative densities and packing fractions of the soil particles and fluid, Eq. \ref{eqn:failradvert} reduces to 

\begin{equation}
\frac{R_f}{R_0} \approx \left(\frac{2}{1-\frac{K_a}{K_0}}\right)^{\frac{1}{2}}.
   \label{eqn:simplefailrad}
\end{equation}
Equation 11 facilitates a prediction of $R_f$ using only two soil properties, $K_a$ and $K_0$, both of which are commonly measured during a geotechnical survey \cite{astmd4767-04}. $K_a$ has an established relationship with the friction angle $\varphi$ as given in Eq. \ref{eqn:coeffactivefail}.

$K_0$, on the other hand, is sometimes written as $K_0=1-\sin{\varphi}$. Using typical friction angles and this equation, sands are predicted to have $K_0\approx0.6-0.7$. However, this equation is generally accepted as only a starting point for many substrates. For sand, it may underestimate this ratio, as the material can overconsolidate \cite{LEP}.  We will include  0.6 in our calculations as a lower limit. And to calculate the full range of possible $\frac{R_f}{R_0}$ we will include the possibility of $K_0$ up to 1, as was measured in a very similar system \cite{winter2010thesis}.

Applying the \it full \rm range of possible $K_a$ and $K_0$ values to Eq. \ref{eqn:simplefailrad} yields $1 < \frac{R_f}{R_0} < 5$ in most conditions (Fig. \ref{fig:model}C). These results demonstrate that soil failure around a contracting cylindrical body is a relatively local effect, and for reductions of $p_i$ to near zero, depth-independent. Equation \ref{eqn:simplefailrad} also does not depend on any soil cohesion terms, indicating that localized substrate failure and fluidization should be possible in both granular and cohesive soils.  

Equation \ref{eqn:simplefailrad} thereby gives a hard prediction for what the size of the fluidized region should be around RoboClam. Further, as long as $t_{in}$ is larger than $t_{crit}$, the size of the region should be fixed. And as argued before, if $t_{in}$ is substantially less than $t_{crit}$, the size of the zone should be smaller since particles will not be able to advect. 

For our particular values of $K_0$ and $K_a$, incorporating uncertainties from the friction angle and $K_0$ we expect then that $1.6 < \frac{R_f}{R_0} < 2.2$, and specifically predict $ \frac{R_f}{R_0} \approx 1.7$ for $K_a=1$ and $\varphi=34^{\circ}$. As it is more straightforward to compare our data to $R_e$, which is the radius of contracted RoboClam, we make a further calculation, transforming $R_0$ into $R_e$.  This gives an expected range of $\frac{R_f}{R_e}$ from 2.2 to 3.1, and a specific prediction $ \frac{R_f}{R_e} \approx 2.4$.

\section{Experimental Results and Discussion}

\begin{figure*}[ht!]
\centering
\includegraphics[width=2\columnwidth]{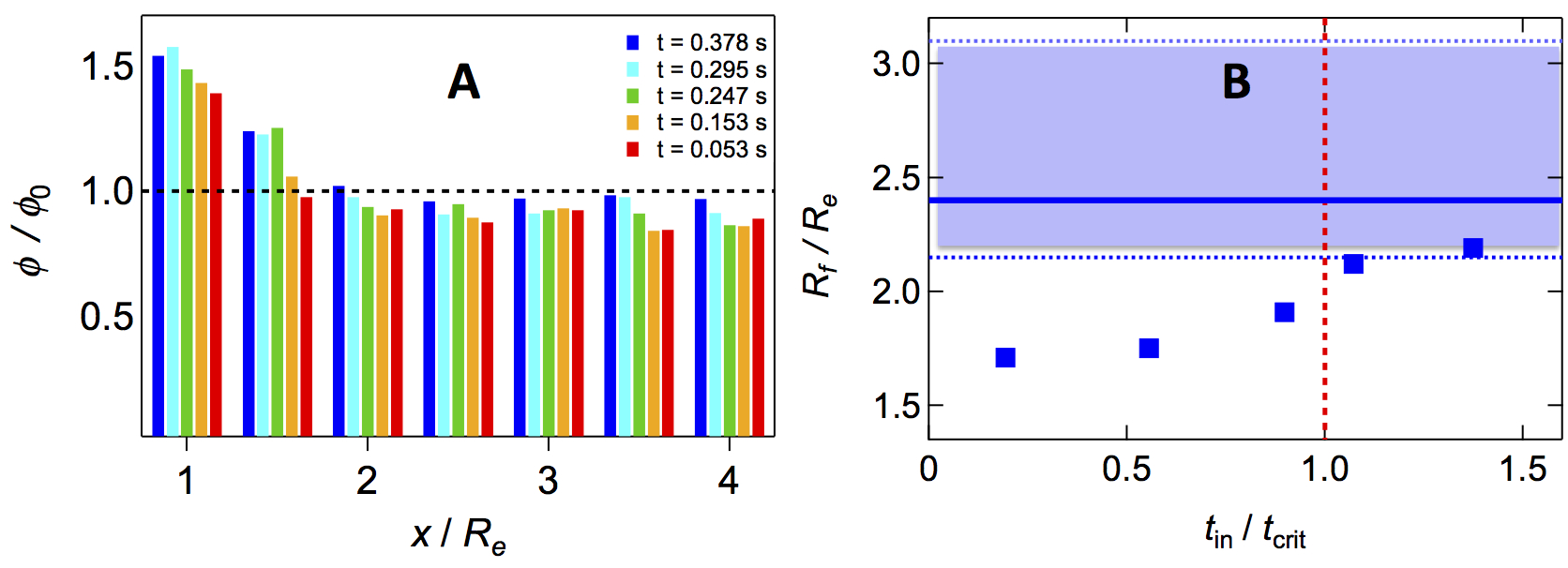}
 \caption{Quantifying local fluidization. (A) Localized fluidization around the end effector. Each histogram corresponds to a different contraction time, and depicts void fraction of the substrate as a function of distance from the end effector. The data are normalized with respect to the fluidization void fraction $\phi_0=$0.41. The fluidization void fraction is marked by the black dashed line; fluidization corresponds to data above the line.  The position of crossover to fluidization yields plot (B), the size of the fluidized zone, for different contraction times, normalized by the critical advection time. Error bars are smaller than the symbol size. The red dashed line corresponds to the ratio of $t_{in}$ to the $t_{crit}$ equalling one. The shaded region bounded by dashed blue lines indicates the predicted range of $R_f/R_e$, with the solid blue line denoting the specific numerical prediction. }
 \label{fig:histogram},
\end{figure*}

We visualize the size of the fluidized regions in the material by looking at the local displacements within the material. We obtain these displacements by the unique identification of particles \cite{Steve} and creation of  particle tracks. We can then create ``speed fields'' by looking at the absolute displacements in the plane of the contraction, interpolated onto a grid. We use absolute displacement, as a particle may be fluidized (move more quickly) without necessarily following the exact vectorial path of  the end effector or its neighbors. In other words, any significant speed, even if against the grain, should necessarily count as a free, fluidized region. In Fig. 5, we display these plots for five different contraction times.  The leftmost plot corresponds to the fastest contractions, and decrease in contraction speed going to the right. The plots are also decorated by predictions for the extent of the fluidized zone (blue lines) and the failure angle (green dashed line). We define a cutoff distance $D$ for the purposes of visualization: particles displacing more than $D$ will be considered in actively moving regions, and have a white color in the graph. For Fig. 5, we have defined this cutoff distance to be one third of the end effector's displacement. We can comfortably adjust this cutoff by about a factor of two in either direction, and still get the same qualitative pictures. Completely static regions are shaded in black, and slow (but moving) regions are in red and yellow.

In Figs. 5A-E we do see the presence of a locally mobile region in all plots. As the contraction time increases, the absolute speed of particles in the region increases, suggesting that the particles are more effectively fluidized. At the shortest time, the particles are mobile, but do not approach the speeds of the longest time. We also see the fluidized zone tends to shrink as the contraction rate is increased ($t_{in}$ is decreased), which aligns with our expectation from fluid dynamical considerations. The apparent size of the zone seems to qualitatively agree with our predictions (blue lines). We also see the presence of the failure wedge in Figs. 5A-E, predicted at $\theta_f=62^{\circ}$. Particles inside the failure wedge that were not advected with the flow are starting to landslide towards the end effector. This wedge is mostly clearly developed for longer contraction times; for the shorter times the elapsed time is not on par with the landslide time of the material ($\approx$ 0.5 s).


 


To explicitly measure where fluidization occurs, we measure the void fraction in the system, by identifying the local neighbors within a 100 pixel radius of each particle. The volume fraction of our undisturbed, randomly packed sample is 0.62. The average area of a sphere in a 2D slice is $A = \frac{2}{3}\pi r^2$, where $r$ is the particle radius \cite{math}. Thus the local packing/void fraction may be inferred by counting the distribution of neighbors within a certain radius, and assuming a random slice. By averaging the void fraction over all vertical positions, we can measure the horizontal extent of the fluidized zone. Fig. 6A shows the void fraction $\phi$ for different contraction times as a function of horizontal distance from the end effector. Defining fluidization as a void fraction of $\phi_0=$0.41, as in \cite{2012expbiorazorclam}, this gives a direct measurement of the size of the fluidized zone. For each contraction time data set, we fit the four (normalized) void fraction vs. position data points closest to the end effector to a polynomial. We can thus measure the extent of the fluidized region by seeing where this polynomial has a value of 1.  We plot the results of this procedure in Fig. 6B. We see that the size of the zone matches the prediction of the mechanical theory for longer contraction times, $\frac{R_f}{R_e}=2.2$. We predicted this ratio to be specifically 2.4, but 2.2 is well in the range of our uncertainty. Interestingly, this measured ratio constrains our value of $K_0$ to indeed be approximately 1. We also see that the number is consistent for contraction times longer than $t_{crit}$, and smaller for shorter times, which aligns with our predictions. 

It is important to underscore that we would not expect a discontinuous ``turn-on" of fluidization at $t_{in}/t_{crit}\approx 1.$ There is no phase transition occurring here, it is simply a competition between particle advection and fluid flow. If the advective motion timescale is larger (more particle inertia), the fluidization is less. However, we never get to a regime where $t_{in}$ is so short that the particles do not advect at all. Thus we see what looks to be a continuous transition. On the other hand, the limiting behavior of this curve might be of future interest, and could be measured with a wider dynamic range of $t_{in}$. This would be an interesting exploration for future studies.


Due to the granular nature and finite size of the system, we also looked into nonaffine motions within the system. Nonaffine motions can be the result of a variety of phenomena, including irreversible rearrangements or force chain breakage. Nonaffine motions point to deviations in the mechanical behavior of the granular material from an ideal viscous, elastic, or viscoelastic medium. In short, the presence of significant nonaffine motion suggests that continuum models are not valid. To measure nonaffine motion, we use the quantity $D^2_{min}$ as we have in previous work [cite]: $D^2_{min,i}=\text{min}\lbrace\sum_j[\Delta\overline d_{ij}(t)-E_i\overline d_{ij}]\rbrace ^2$. $D^2_{min,i}$ quantifies the nonaffine motion of $j$ particles in the neighborhood around a given particle $i$ after removing the averaged linear response to the strain, given by tensor $E_i$; a larger $D^2_{min}$ indicates more nonaffine motion. The vector $\overline d_{ij}$ is the relative position of $i$ and $j$,  $\Delta\overline d_{ij}$ is the relative displacement. 

We have compared the nonaffine motion to the total displacements, and find no trends with contraction time or position. Further, nonaffine motion accounts for less than 5 percent of displacement in all trials and frames. This might be surprising, considering this is a granular system to begin with, where rearrangements and force chain breakages are significant events. It also suggests that our continuum model is valid for use, despite the fact that the diameter of our grains is only a factor of 5 less than the size of the end effector. But upon reflection, this is exactly what we should expect, as the system is not truly granular. The fluidized region has no force chains to break, and the particles advect with the fluid. The particles outside the failure wedge remain stationary. Only in the late ``landslide'' behavior should nonaffine motions be in any way significant, but this is not of interest for the model.

\section{Conclusions}

In conclusion, we have shown that a previously developed mechanical model for \it E. directus \rm captures the fluidization dynamics of RoboClam within a 3D granular bed.  Specifically, it is shown that the size of the fluidized region is the size we expect it to be based on this model: roughly the size of the end effector itself. What can be tested in future work is further variation of soil, fluid, and effector parameters. The mathematical model can incorporate these variations, it is yet to be determined if the model breaks down at some point. 

We have also shown if the contraction time is too short, the fluidized region will become smaller, because the particles will fluidize less effectively. We have shown that as the contraction time increases the fluidized region becomes larger. While this points to maximizing the contraction time as one design goal, it is not the only timescale: future experiments must also look at the interplay between the timescales of fluidization, settling, and landslides.

We also have seen the result that nonaffine motion is actually quite insubstantial in this system. This is somewhat counterintuitive not only because it is a granular system to begin with, but also because the length scales of the particles are comparable to the end effector size. Since it is a granular system, one expects nonaffine effects to become important for dynamics - however, if the system is always fluidized, this just may be unimportant. The result ultimately supports the use of this continuum model for this system; since deviations from the average are small, a continuum model works well even with large particles. Where these motions may become more important is in the downward digging motion itself: while the grains on the side are fluidized, the grains below are still packed together, a topic for future exploration.

\begin{acknowledgements}

We acknowledge support from Bluefin Robotics, and U.S. DTRA under
Grant No. HDTRA1-10-0021. We thank Robin L.H. Deits for previous work. We thank Don Martin for technical support.

\end{acknowledgements}
\bibliography{Roborefs}

\begin{thebibliography}{29}
\bibitem [{mat()}]{math}%
  \BibitemOpen
  \href@noop {} {}\bibinfo {note} {We are randomly sampling cross-sectional
  areas, and so are interested in the average area. For any object, the average
  cross-sectional area over some depth $Z$ is $\langle
  A\rangle_z=\frac{1}{Z}\int_0^Z A(z)dz$. If $Z$ corresponds to the diameter of
  the sphere, then the integral is simply the volume, and $\langle
  A\rangle_z=\frac{1}{2r}\frac{4}{3}\pi r^3 = \frac{2}{3}\pi r^2$.}\BibitemShut
  {Stop}%
\end{thebibliography}

\end{document}